\documentclass[superscriptaddress,aps,twocolumn,prb,amsmath,amssymb,longbibliography]{revtex4-1}
\usepackage[varg]{txfonts}
\usepackage{color}
\usepackage{hyperref}

\usepackage{graphicx}
\usepackage{bm}

\def\XXint#1#2#3{{\setbox0=\hbox{$#1{#2#3}{\int}$}
    \vcenter{\hbox{$#2#3$}}\kern-.5\wd0}}

\def\bA{{\bf A}}
\def\bB{{\bf B}}
\def\bE{{\bf E}}
\def\bj{{\bf j}}

\def\bk{{\bf k}}
\def\br{{\bf r}}
\def\bu{{\bf u}}

\def\bnabla{{\bm \nabla}}

\def\be{\begin{equation}}
\def\ee{\end{equation}}

\def\bi{\begin{itemize}}
	\def\ei{\end{itemize}}
\def\bn{\begin{enumerate}}
	\def\en{\end{enumerate}}
\def\bea{\begin{eqnarray}}
\def\eea{\end{eqnarray}}
\newcommand{\bpm}{\begin{pmatrix}}
	\newcommand{\epm}{\end{pmatrix}}

\def\ba{\begin{array}}
	\def\ea{\end{array}}
\def\bd{\begin{displaymath}}
\def\ed{\end{displaymath}}

\renewcommand{\imath}{\hspace{1pt}\mathrm{i}\hspace{1pt}}

\begin{document}

\title{Dynamo Effect and Turbulence in Hydrodynamic Weyl Metals}

\author{Victor~Galitski}
\affiliation{Joint Quantum Institute, Department of Physics,
University of Maryland, College Park, MD 20742-4111}

\author{Mehdi Kargarian}
\affiliation{Department of Physics, Sharif University of Technology, Tehran 14588-89694, Iran}

\author{Sergey Syzranov}
\affiliation{Physics Department, University of California, Santa Cruz, CA 95064, USA}
\affiliation{Joint Quantum Institute, Department of Physics, University of Maryland, College Park, MD 20742-4111}

\begin{abstract}
The dynamo effect is a class of macroscopic phenomena responsible for generation and maintaining magnetic fields in astrophysical bodies. It hinges on hydrodynamic three-dimensional motion of conducting gases and plasmas that achieve high hydrodynamic and/or magnetic Reynolds  numbers due to large length scales involved. The existing laboratory experiments modeling dynamos are challenging and involve large apparatuses containing conducting fluids subject to fast helical flows. Here we propose that electronic solid-state materials -- in particular, hydrodynamic metals -- may serve as an alternative platform to observe some aspects of the dynamo effect. Motivated by recent experimental developments,  this paper focuses on hydrodynamic Weyl semimetals,
where the dominant scattering mechanism is due to interactions. We derive Navier-Stokes equations along with equations of magneto-hydrodynamics that describe transport of  Weyl electron-hole plasma appropriate in this regime. We estimate the hydrodynamic  and magnetic Reynolds numbers for this system. The latter is  a key figure of merit of the dynamo mechanism.  We show that it can be relatively large to enable observation of the dynamo-induced magnetic field bootstrap in experiment. Finally, we generalize the simplest dynamo instability model -- Ponomarenko dynamo -- to the case of a hydrodynamic Weyl semimetal and show that the chiral anomaly term reduces the threshold magnetic Reynolds number for the dynamo instability.
\end{abstract}

\maketitle

The dynamo effect is a beautiful astrophysical phenomenon, first proposed by Larmor in 1919~\cite{Larmor},  that is believed to be responsible for generating and sustaining magnetic fields in galaxies, stars and planets including the Sun and the Earth~\cite{RSZ}. There exist a large variety of different dynamo mechanisms~\cite{Gilbert,Gilbert1,RSZ}
that all share the same key ingredient -- hydrodynamic motion of an electrically conducting  gas, fluid or plasma.
The dynamo theory deals with the hydrodynamic motion 
of a conductive medium focussing on the possibility
of self-generating and self-sustaining magnetic fields, whose presence has been observed in astrophysical bodies.

As detailed below, the underlying equations of  the theory are the Navier-Stokes equations, describing the hydrodynamic motion of the medium, coupled to the Maxwell equations of electromagnetism.  In the non-relativistic limit, they give rise to equations of magneto-hydrodynamics (MHD). These are complicated non-linear equations, and their exact solutions represent a great challenge. However, both the solutions of simplified MHD models [e.g., kinematic dynamos, with predetermined velocity fields ${\bf u}({\bf r},t)$] and qualitative arguments~\cite{RSZ}
suggest that
the dynamo action is possible
when the terms enhancing the magnetic field
[e.g. the induction term, ${\bm \nabla} \times ({\bf u} \times {\bf B})$] overwhelm the magnetic diffusion term, $\eta_m \Delta {\bf B}$ (where  $\eta_m = c^2/4 \pi \sigma$ where $c$ is the speed of light and $\sigma$ is the conductivity of the medium), which tend
to suppress the self-generation.
The respective figure of merit is the {\em magnetic Reynolds number}~\cite{LL6}
\begin{equation}
\label{Rm}
R_m = \frac{u L}{\eta_m} = u L \frac{4 \pi \sigma}{c^2},
\end{equation}
where $L$ is the characteristic system size and $u$ is the typical velocity of the medium. The threshold value for a dynamo action to commence
(usually lying in in the range $R_m^{\rm(cr)} \sim 10 - 100$, with $R_m^{\rm(cr)} \approx 17.7$ for the simplest Ponomarenko dynamo~\cite{Ponomarenko} discussed below) depends on system's geometry  and is rarely known exactly.
It is clear, however, that 
the larger $R_m$, the more likely and more effective the dynamo action. The  conductivity of astrophysical media vary greatly from $10^{-11}{\rm Sm}^{-1}$ for interstellar plasma to $10^{3}{\rm Sm}^{-1}$ for the solar convection shell and $10^{5}{\rm Sm}^{-1}$ for the Earth's core, but in all of these cases the large magnetic diffusion coefficient is compensated by literally astronomical distances
resulting in large magnetic Reynolds numbers, however small the conductivities are. 
By contrast, laboratory dynamo experiments~\cite{LathropPT} deal naturally limited system size
and use the conductivity and the flow velocities as the only potentially tunable parameters.

Apart from large magnetic Reynolds numbers $R_m \gg 1$,
the emergence of a dynamo requires 
a number of other conditions that need to be met.
In particular, certain ``no-go theorems''~\cite{Jones}
have to be overcome, such as the impossibility of a two-dimensional dynamo effect or that in a planar three-dimensional flow
(i.e., with one vanishing component of velocity). Finally, it is known 
the dynamo action is greatly helped by the helicity flow,
which may arise either due to the geometry of an imposed flow or due to turbulence. The latter
is possible if the second figure of merit, the {\em hydrodynamic Reynolds number}
\begin{equation}
\label{R}
R = \frac{u L}{\nu},
\end{equation}
where $\nu$ is the kinematic viscosity. Separating both the velocity and magnetic
 field into a mean-field and fluctuating component - 
 ${\bm u} = \overline{\bm u} + \delta {\bm u}$ and ${\bm B} = \overline{\bm B} + \delta {\bm B}$, 
and averaging over the small-scale fluctuations results in the Krause-R{\"a}dler equations~\cite{KR,Parker} of mean-field MHD, which in the simplest case of isotropic turbulence is given by
\begin{equation}
\label{KR}
\frac{\partial \overline{\bf B}}{\partial t} = {\bm \nabla} \times (\overline{\bm u} \times \overline{\bf B}) + {\bm \nabla}  \times (\alpha  \overline{\bf B}) + \xi \Delta \overline{\bf B},
\end{equation}
where the second term in the right-hand-side is the ``new'' helicity term allowed in turbulent MHD ($\alpha$-effect). If the velocity field is stationary, Eq.~(\ref{KR}) or a similar MHD equation without helicity for non-turbulent flows becomes an eigenvalue problem for the magnetic field growth ${\bf B}({\bf r},t) \propto {\bf B}({\bf r})  e^{\gamma t}$. The existence of
exponentially growing components (${\rm Re}\, \gamma >0$)
indicates an instability towards a self-generating magnetic field (were the imaginary part ${\rm Im}\, \gamma >0$ leads to the field oscillations,
which have been suggested~\onlinecite{GS} by one of the authors to lead, e.g., to periodic cycles of solar magnetic activity).

Apart from the astrophysical context, there has been a tremendous interest in testing the predictions of dynamo theory and  modeling a planetary-like or solar-like dynamo action in the laboratory~\cite{LathropPT,DynamoExp,DynamoExp1,DynamoExp2}. Several impressive laboratory experiments have been carried out and are currently under way that involve setting in motion a liquid metal -- sodium or gallium -- with the goal to achieve large Reynolds numbers to enable the dynamo mechanism. As obvious from Eqs.~(\ref{Rm}) and (\ref{R}), this leads to the challenge of ultra-fast mechanical stirring or rotating the liquid metal. 

Here we propose that electronic solid-state systems may provide an alternative platform for observing  magnetohydrodynamic effects.
Firstly, we list several necessary conditions of the dynamo effect in an electronic system:
(i)~Transport in the electron liquid should be governed by hydrodynamics, i.e. the primary momentum relaxation mechanism
should be electron-electron collisions
rather than impurity scattering. (ii)~The system and the flow must be essentially three-dimensional. (iii)~Large magnetic $R_m\gg 1$
and/or hydrodynamic $R\gg 1$ Reynolds numbers are required. 

Hydrodynamic transport in solid state [condition (i)] has been a subject of intense recent studies~\cite{Kivelson_hydro,Hartnoll_Lucas,SDS_hydro,Balents_SYKs,Sachdev_book2},
both theoretical and experimental. On the experimental side, two widely studied platforms for hydrodynamic phenomena
are graphene~\cite{GrapheneHydro} and Weyl semimetals (WSMs)~\cite{Felser1,Felser2,Felser3}. Graphene, however, 
violates a ``no-go dynamo theorem'' -  condition (ii) requiring 3D flows - and is thus 
of no relevance to the dynamo effect.

In what follows,
we focus on undoped or weakly doped Weyl semimetals.  We note that 
in systems with the power-law quasiparticle dispersion $\epsilon({\bf p}) \propto \left| {\bf p} \right|^\beta$ with $\beta\leqslant 1$ the creation of electron-hole pairs is suppressed~\cite{FosterAleiner},
because the energy and momentum conservation laws cannot be satisfied simultaneously for lowest-order
processes.
Weyl systems ($\beta=1$) may, therefore, often be considered as electron-hole
plasma with a linear particle dispersion.

A WSM generically has an even number of nodes, according to the fermion-doubling theorem~\cite{NielsenNinomiya}, and electrons and holes near different nodes often behave as independent liquids. However, simultaneous application of external electric $\bE$ and magnetic $\bB$ fields results in the quasiparticle transfer from one node to another (chiral anomaly\cite{Adler:anomaly,BelJackiw:anomaly,SonSpivak:anomaly,Burkov:review,Burkov:AnomalyDiffusive}). 
For simplicity, we assume in this paper that (a)~the system has only two nodes, labeled by $L$ and $R$,
with the same quasiparticle dispersion, (b)~the entire system is being kept at
a constant temperature $T$ and (c)~the intranodal equilibration processes are significantly faster than the
 internodal particle-transfer processes.
This allows one to define the chemical potentials $\mu_\alpha$ near each node $\alpha=L, R$
and the hydrodynamic velocity $\bu$ of the Weyl fluid. The distribution function of 
the linearly-dispersing quasiparticless near each node in the absence of electromagnetic fields is given by~\cite{Narozhny:grapheneHreview} 	$f_\alpha(\bk)=\left\{\exp\left[\gamma(\bu)\left(\pm v_F|\bk|-\mu_\alpha-\bu\cdot\bk\right)/T\right]+1\right\}^{-1}$, 
	where ``+'' and ``-'' refer, respectively, to the conduction and valence bands,
	and $\gamma(\bu)=\left(1-u^2/v_F^2\right)^\frac{1}{2}$.

The dynamics of charge densities $\rho_\alpha$ near node $\alpha$, where $\alpha=L,R$,
 are described by the continuity equations
\begin{align}
	\partial_t \rho_\alpha+{\bm \nabla\cdot {\bf j}_{\alpha}}-\chi_{\alpha}\frac{g e^3 }{4\pi^2 \hbar^2 c}
	\bE\cdot\bB
	+\frac{\rho_\alpha-\rho_{\bar{\alpha}}}{\tau_{in}}=0,
	\label{ChargeContinuity}
\end{align}
where $\chi_{L}=-1$ and $\chi_R=+1$ are the ``chiralities'' of quasiparticles 
near nodes $L$ and $R$ and $g$ accounts
for spin and possibly additional valley degeneracy; $\bar{\alpha}$
labels the node other than $\alpha$; hereinafter $e=-|e|$.
The first two terms in Eq.~(\ref{ChargeContinuity}) match the usual continuity equation for a liquid
with density $\rho_{\alpha}$; the third term ($\propto \bE\cdot\bB$) accounts~\cite{Burkov:review,Burkov:AnomalyDiffusive}
 for the change of the electron
concentration at node $\alpha$ due to the chiral anomaly; and
the last term in Eq.~(\ref{ChargeContinuity}) describes internodal scattering, e.g., due to short-range-correlated
quenched disorder, with the internodal scattering time
$\tau_{in}$.
The electric currents $\bj_{L,R}$ of the charge carriers near the two nodes are given by
\begin{align}
	\bj_\alpha = \sum_\beta\sigma_{\alpha\beta}
	\left[\bE+\frac{1}{c}\bu\times\bB-\frac{1}{e}{\bm \nabla}\mu_\beta\right]
	-\chi_\alpha\frac{ge^2}{4\pi^2\hbar^2c}\bB\mu_\alpha,
	\label{Currents}
\end{align}
where $\alpha,\beta=L,R$;  $\mu_\alpha$ is the chemical potential near node $\alpha$, and ${\bf u}$ is the hydrodynamic velocity of the Weyl fluid.
In this paper we assume that the imbalance of the chemical potentials between the nodes, if any, is small  $|\mu_L-\mu_R|\ll |\mu_{L,R}|,T$.
The diagonal components $\sigma_{LL}=\sigma_{RR}$
of the conductivity tensor $\sigma_{\alpha\beta}$ describe the response of charge carriers near each
node to the electromagnetic field; the off-diagonal entries $\sigma_{LR}=\sigma_{RL}$ account
for the drag of the quasiparticles near each node
by the current near the other node. The last term in Eq.~(\ref{Currents}) describes the
chiral magnetic effect~\cite{Kharzheev:CME,Kharzheev:CMTlextures},
the generation of the charge current by an external magnetic field in the system in the
presence of chirality imbalance,
$\mu_L-\mu_R\neq 0$. 

Equations (\ref{ChargeContinuity})-(\ref{Currents}), together with the relations~\cite{RodionovSyzranov:impurityWeyl}
\begin{align}
	\rho_{R,L}=ge\frac{\mu_{R,L}^3+\pi^2\mu_{R,L}T^2}{6\pi^2 v_F^3 \hbar^3}
	\label{RhoMuRelation}	
\end{align}
for the charge density at node $\alpha$
and with Maxwell equations, which involve the total charge density $\rho=\rho_L+\rho_R$
and the current $\bj=\bj_L+\bj_R$,
constitute a closed system of equations which describes charge and current dynamics 
of the electron liquid in a WSM which moves with velocity $\bu$ in an external electromagnetic
field. The motion of such a liquid may be generated by the electromagnetic field,  the temperature
and chemical potential gradients, or even fast mechanical rotation of the sample.

To determine self-consistently the velocity field $\bu$
(which in practice is a tremendously difficult problem), the system of Eqs.~(\ref{ChargeContinuity})-(\ref{RhoMuRelation}) has to be complemented by the  Navier-Stokes equation (derived in Supplemental Material~\cite{suppl_dynamo})
\begin{align}
\frac{w_\alpha}{v_{F}^2}
\left(\frac{\partial}{\partial t}+\bu\cdot\bnabla \right)\bu=
-\bnabla P_\alpha-\frac{\bu}{v_F^2}\frac{\partial P_\alpha}{\partial t}
+\rho_\alpha\bE+\frac{1}{c}\,\bj_\alpha\times\bB
\nonumber\\
+\frac{\bu}{3}\left(\frac{\partial\varepsilon}{\partial\rho}\right)_\alpha
\left(\chi_\alpha\frac{ge^3}{h^2c}\bE\cdot\bB-\frac{\rho_\alpha-{\rho}_{\bar{\alpha}}}{\tau_{\text{in}}}\right)
+\eta \bnabla^2\bu + \zeta \bnabla\left(\bnabla\cdot\bu\right),
\label{NS}
\end{align}
where $w_\alpha=\varepsilon_\alpha+P_\alpha$ is the the enthalpy 
of the charge carriers near node $\alpha$ per unit volume,
with~\cite{Gorbar:hydroEq}
\begin{align}
	\varepsilon_\alpha\approx g\frac{7\pi^4 T^4+30\pi^2\mu_\alpha^2 T^2+15\mu_\alpha^4}{120\pi^2v_F^3\hbar^3}
	\label{EpsilonMuRelation}
\end{align}
 and $P_\alpha\approx\frac{\varepsilon_\alpha}{3}$
being, respectively, the contributions of node $\alpha$ to the internal energy and pressure; the current $\bj_\alpha$ is given by
Eq.~(\ref{Currents}); $\eta$ and $\zeta$
are the shear and the bulk viscosities; the term $\propto\left(\frac{\partial\varepsilon}{\partial\rho}\right)$
accounts for the change of the energy and pressure of the Weyl liquid near node $\alpha$ due to the internodal scattering
and the chiral anomaly, where
$\left(\frac{\partial\varepsilon}{\partial\rho}\right)_\alpha=
\frac{3\mu_\alpha}{e}\frac{\mu_\alpha^2+\pi^2 T^2}{3\mu_\alpha^2+\pi^2 T^2}$ for the case 
of an isothermal flow considered in this paper [see Supplemental Material~\cite{suppl_dynamo} for the discussion
of the assumptions about thermalisation].

In this paper, we neglect the so-called chiral vortical effect~\cite{Gorbar:hydroEq}, i.e. contributions to the current from the interplay of global rotations of the system
and chirality imbalance ($\mu_L-\mu_R\neq0$). 
In the Navier-Stokes equation~(\ref{NS}) we also neglect terms of higher orders in $u^2/v_F^2$. 
Equations~(\ref{ChargeContinuity})-(\ref{NS}), together with the Maxwell's equations and
the equations of state, in the form of Eq.~(\ref{EpsilonMuRelation}) and $P_\alpha=\frac{\varepsilon_\alpha}{3}$, constitute a closed system of equations describing
the dynamics of the electromagnetic fields and the electron liquid in a WSM.

Using Eqs.~(\ref{Currents}),
together with the Maxwell's equations $\bnabla \times\bE=-\frac{1}{c}\frac{\partial\bB}{\partial t}$
and $\bj_L+\bj_R\equiv\bj=\frac{c}{4\pi}\bnabla\times \bB$, where we neglected the displacement current 
under the assumption of a quasi-stationary flow, we arrive at the equation for the dynamics of the magnetic field:
\begin{align} \label{Eq:dyn-helical}
	\frac{\partial \bB}{\partial t} = {\bm \nabla} \times ({\bm u} \times{\bf B})
	+\frac{c^2}{4\pi\sigma}\nabla^2 \bB
	+\frac{ge^2}{4\pi^2\hbar^2\sigma}{\bm\nabla}\times\left[(\mu_L-\mu_R)\bB\right],
\end{align}
where $\sigma=2\sigma_{LL}+2\sigma_{LR}$ is the conductivity of the WSM and we have taken into account
that the quasiparticles have the same dispersion near the two nodes. 
{Apart from solid-state WSMs, an equation of the form 
	(\ref{Eq:dyn-helical}) with phenomenologically introduced coefficients describes
	the dynamics of ultrarelativistic chiral particles~\cite{Yamamoto:chiralLiquid}.
}

Equation~(\ref{Eq:dyn-helical}) indicates that
Weyl liquids allow for the helicity term for macroscopic fields without turbulence,
in contrast with the conventional $\alpha$-dynamo of Krause and R{\"a}dler~\cite{KR}. 
However, it can only appear in the presence of an already existing field, and while, as shown below, it can  further enhance magnetic field ``bootstrap,''  it can not lead to generation of the field in and by itself if there is no seed field to begin with. For that, the magnetic Reynolds number (\ref{Rm}), $R_m$, has to be large enough, as discussed in the introduction. 

 To estimate, $R_m$, we use the equation for the Coulomb-interaction dominated conductivity of a Weyl
semimetal~\cite{HosurVishwanath}
\begin{equation}
\label{sigma} 
\sigma \sim \frac{e^2}{\hbar} \frac{k_B T}{\hbar v_F} \frac{1}{\alpha^2},
\end{equation}
where the Weyl's ``fine-structure constant'' is $\alpha = e^2/(\hbar v_F \varkappa)$ and $\varkappa$ is the dielectric constant, which crucially may be rather large. While  Eq.~(\ref{sigma}) has been derived neglecting screening effects~\cite{HosurVishwanath}, it should be adequate for estimates. For these purposes, we have also dropped logarithmic renormalisation factors. 

\begin{figure}[htbp]
	\centering
	\includegraphics[width=0.5\textwidth]{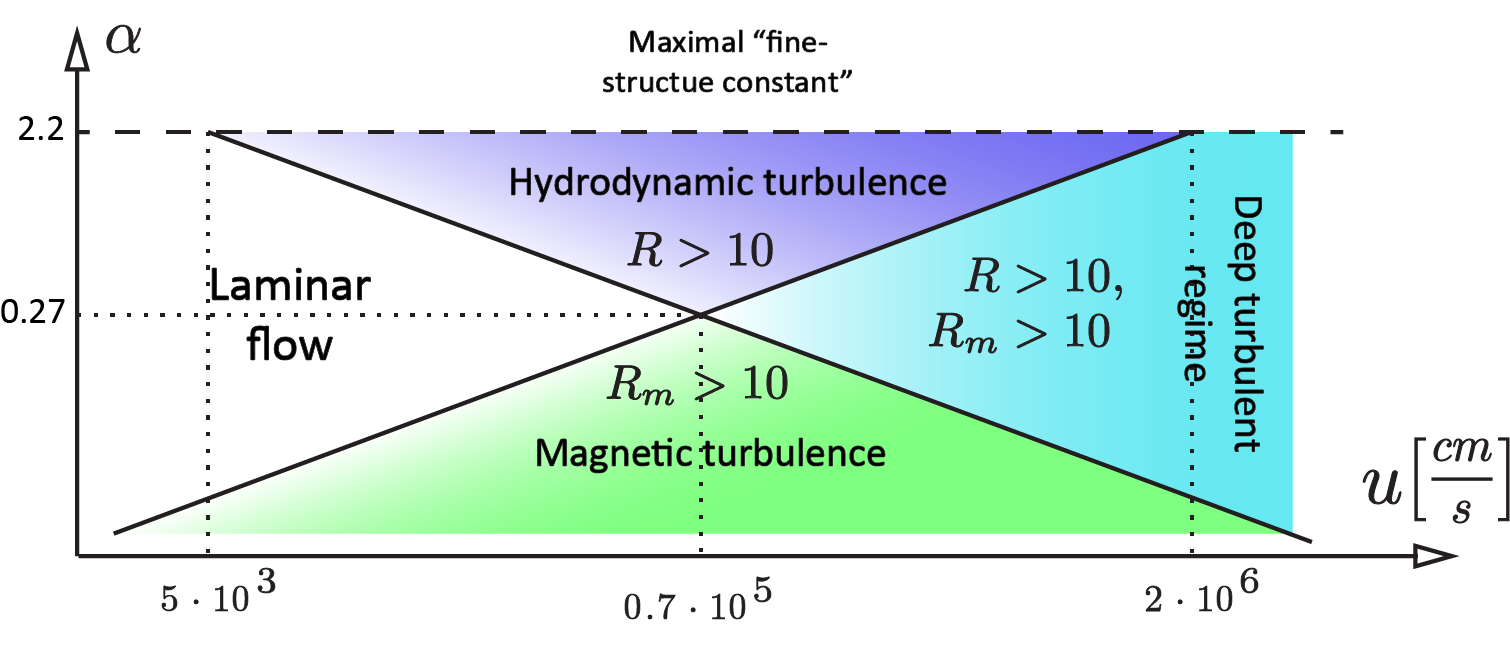}
	\caption{(Colour online) Flow regimes for the electron liquid in a Weyl semimetal on the diagram 
	``fine-structure constant'' $\alpha=\frac{e^2}{\varkappa\hbar v_F}$ vs. flow velocity $u$ (log-log scale)
	for the room temperature $T=T_{\text{room}}=300K$ and the Fermi velocity $v_F=10^8\frac{cm}{s}$.
	The maximum value of the ``fine-structure constant'' is $\alpha_{max}=\frac{e^2}{\hbar v_F}\approx 2.2$.
		\label{RegimesPlot}
    }
\end{figure}

Let us emphasise that the dynamo effect 
is a {\em macroscopic classical phenomenon.} 
The effect if favoured by large system sizes $L$, which lead to large $R_m$.
In experiments with solid-state systems the size $L$ is rather limited,
with centimetre-size samples being at the upper end of the range accessible for WSMs.
Since the effect is not sensitive to quantum interference effects, higher 
temperatures $T$ are much preferable to maximize $R_m$;
the room temperature, $T_{\rm room}$, thus represents a reasonable comparison scale.
We emphasise that even at room temperature Weyl semimetals are not Maxwell gases and quantum statistics and quantum nature of the electron-electron scattering are important,
but quantum coherence is not essential for the dynamo effect. Using these length and temperature scales,  we obtain the following estimate for the main figure of merit in the dynamo theory:
\begin{align}
\label{RmWeyl} 
R_m \sim 
\frac{1}{\alpha^2}\frac{e^2}{\hbar}\frac{4\pi k_B T}{\hbar v_F c^2 }uL
\sim \frac{10^{-6}}{\alpha^2} 
\left( \frac{T}{T_{\rm room}} \right)\times u \left[{\rm \frac{cm}{s}}\right]\, L [{\rm cm}],
\end{align}
where $u$ is the typical velocity of the flow.

Now, we turn to estimates of the hydrodynamic Reynolds number \ref{R}. The viscosity of the quasiparticles in a Weyl semimetal at temperature $T$ may be estimated as $\eta\sim n(T)T\tau_{\text{rel}}$, where $n(T)$ is the concentration of the thermally excited quasiparticles
and $\tau_{\text{rel}}\sim \hbar(\alpha^2 k_B T)^{-1}$ is the momentum relaxation  time. Note that this result follows from second-order perturbation theory in Coulomb interaction and neglects screening effects. This leads to
\begin{align}
	\eta\sim\frac{(k_BT)^3}{\alpha^2\hbar^2 v_F^3}.
\end{align}

The motion of a Weyl-semimetal liquid is turbulent in the hydrodynamic sense when the term $\frac{w}{v_F^2}({{\bf u}\cdot\boldsymbol\nabla}){\bf u}$  in the Navier-Stokes equation (\ref{NS}) dominates the dissipative terms $\sim \eta\nabla^2{\bf u}$ that come from 
the viscosity of the Weyl fluid. This yields the following estimate 
\begin{align}
	R=\frac{wuL}{\eta v_F^2}\sim\alpha^2\frac{k_B T}{\hbar}\frac{uL}{v_F^2} 
	\sim 4\alpha^2 10^{-3}\left( \frac{T}{T_{\rm room}} \right)
	\times u \left[{\rm \frac{cm}{s}}\right]\, L [{\rm cm}],
	\label{Restimate}
\end{align}
where we have used the estimate $w\approx\frac{7g\pi^2T^4}{90\hbar^3v_F^3}\sim\frac{(k_B T)^4}{\hbar^3v_F^3}$ for the specific enthalpy
at high temperatures.

We note in this context that the viscosity of a Fermi liquid at temperature $T$ may be estimated as $\eta \sim{\varepsilon_F^5}/({T^2 \hbar^2 v_F^3})$, where $\varepsilon_F$ and $v_F$ are the Fermi energy and velocity, respectively. Because the hydrodynamic Reynolds number $R\sim \frac{T^2}{v_F^2\varepsilon_F\hbar}$ gets rapidly suppressed with increasing the Fermi energy $\varepsilon_F$, topological semimetals are indeed a favourable platform for achieving electronic turbulence as compared to ``conventional'' hydrodynamic metals. 

Naturally, the geometry and the 
magnitude of the velocity field $u$ much depends on the mechanism to stir up hydrodynamic motion and follows from the solution of the Navier-Stoker equations, which is a challenging task in most cases. Furthermore, since observation of a phenomenon of this kind has never been attempted in solid-state materials, it is not clear at the moment what experimental technique would be the most efficient to achieve high hydrodynamic flows -- pulsed fields, crossed electric and magnetic fields or just a rapid rotation of the sample are all possibilities to consider. 
While below we consider in detail 
one of the standard and simplest dynamo models,
we emphasise immediately that the estimates (\ref{RmWeyl}) and (\ref{Restimate})
are not prohibitive; and it is conceivable that relatively large magnetic Reynolds numbers,
necessary for the dynamo to commence,
are achievable
for realistic flow velocities with $u$ of order one kilometre/second or greater
(especially considering that the dielectric constant may be as high as $\varkappa\sim 50$ in WSMs),
cf. Fig.~\ref{RegimesPlot}.

Now, we discuss a specific model of dynamo effect -- the so-called kinematic Ponomarenko dynamo~\cite{Ponomarenko,Jones}, with an eye on how the terms in MHD equations, descending from the chiral anomaly, change the effect. The Ponomarenko dynamo does not necessarily represent the most experimentally realistic setup, but it does represent the simplest textbook model, which contains the key qualitative features of a dynamo mechanism  and is amenable to analytical analysis.  

 In order for a dynamo action to occur, the magnetic Reynolds number must exceed a critical value $R^{c}_{m}$ \cite{plasma}. The purpose of the calculation below is to obtain the dependence of the critical Reynolds number, $R^{c}_{m}$, on the helicity term. For simplicity, we 
neglect the time dependence of the chemical-potential difference $\mu_L-\mu_R$ on the times we consider.

We re-write Eq.~(\ref{Eq:dyn-helical}) as
\begin{align} \label{Eq:dyn-helical2}
	\frac{\partial \bB}{\partial t} = {\bm \nabla} \times ({\bm u} \times{\bf B})
	+\frac{c^2}{4\pi\sigma}\nabla^2 \bB+\xi{\bm\nabla}\times\bB,
\end{align}
where $\xi=ge^2(\mu_L-\mu_R)/(4\pi^2\hbar^2\sigma)$. We consider a cylindrical geometry of the sample with a flow field $\mathbf{u}=(0,r\Omega,u_{0})$, where $\Omega$ and $u_{0}$ are constants, for $r\leq a$, and $\mathbf{u}=0$ for $r>a$ \cite{plasma}. Plugging the ansatz $\mathbf{B}(r,\theta,z,t)=\mathbf{B}(r) e^{i(n\theta-kz)+\gamma t}$ into (\ref{Eq:dyn-helical}), the components of the magnetic field $B_{\pm}=B_{r}\pm iB_{\theta}$ satisfy the equations
\begin{align} \label{dynamoEq1} 
y^2 B''_{\pm}+y B'_{\pm}=\left[q^2y^2+(n\pm1)^2\right]B_{\pm}\nonumber 
\\-\delta\left[nyB'_{\mp}\mp n(n\mp1)B_{\mp}\pm k^2a^2y^2B_{\pm}\mp q^2y^2B_r  \right] \end{align}     
for $y=r/a\leq1$ and  
\begin{align}  \label{dynamoEq2} 
y^2 B''_{\pm}+y B'_{\pm}=\left[s^2y^2+(n\pm1)^2\right]B_{\pm} \end{align} 
for $y>1$, where $B'_{\pm}$ ($B''_{\pm}$) is the first (second) derivative with respect to $y$; $\delta=4\pi\sigma \xi/kc^2$, $q^2=k^2a^2+\gamma\tau_{R}+i(n\Omega-ku_{0})$, $s^2=k^2a^2+\gamma\tau_{R}$, where $\tau_{R}=4\pi\sigma a^2/c^2$ is the time scale of the magnetic field diffusion. 

\begin{figure}[htbp]
	\centering
	\includegraphics[width=0.45\textwidth]{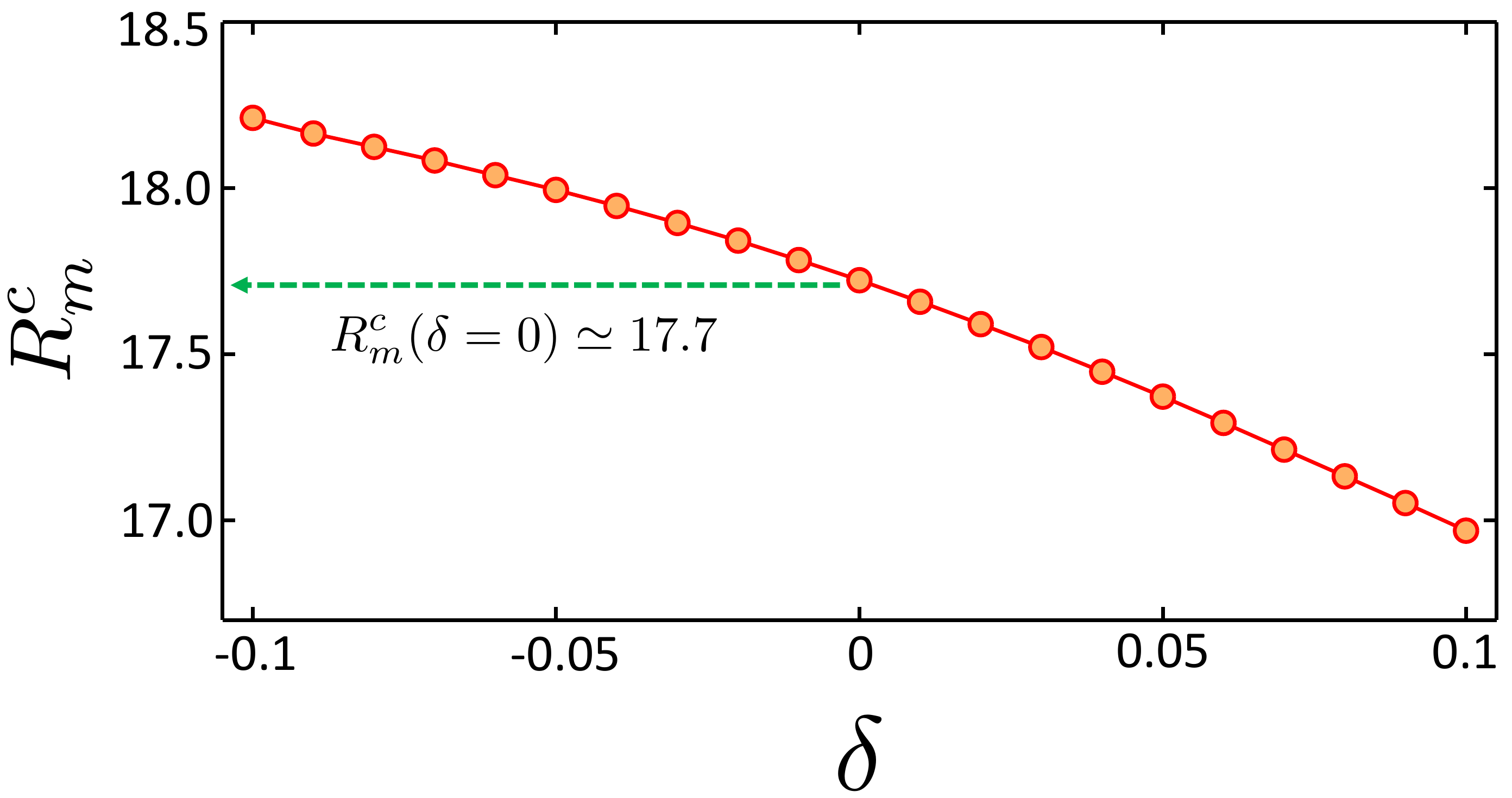}
	\caption{(Colour online) The critical magnetic Reynold number $R_{m}^{c}$ of the $n=1$ kinematic Ponomarenko dynamo 
	as a function of the helicity parameter $\delta=4\pi\sigma \xi/kc^2$, with $k$ being the wave-vector of the dynamo instability. 	$R^{c}_{m}\simeq 17.7$ is the critical value for the dynamo in the absence of helicity. We note that a self-exciting dynamo will always correspond to the chirality with a lower critical Reynolds number. The chiral anomaly, thus, always aids the dynamo effect.}
    \label{critical_Rm}
\end{figure}


For each mode $n$, the magnetic field starts to grow exponentially when $\mathrm{Re}(\gamma)>0$, which occurs if the magnetic Reynolds number exceeds a critical value $R_{m}^{c}$. In the absence of helicity ($\delta=0$), Eq.~(\ref{Eq:dyn-helical2}) reduces to the conventional dynamo equation and the $n=0$ mode is not excited for an arbitrary intensity of the flow \cite{plasma}. 
For non-zero helicity, we solved the inhomogeneous equations (\ref{dynamoEq1}) and (\ref{dynamoEq2}) with appropriate boundary conditions imposed \cite{suppl_dynamo} to obtain the dispersion relation for the dynamo mode.
The obtained values of $R_{m}^{c}$ {for a dynamo with $n=1$ and a particular direction of wavevector $\bk$ (the $z$ axis)}
are shown in Fig.~\ref{critical_Rm}.
The $n=1$ mode is the leading mode, where the dynamo action commences first, and for which the critical magnetic Reynold number is the smallest and potentially within reach for actual Weyl systems. In the absence of helicity (i.e., if $\delta=0$), it is known to be $R^{c}_m\simeq17.7$ \cite{plasma}.  Interestingly enough, the helicity $\delta>0$ reduces the critical value of the magnetic Reynold number for the $n=1$ mode and helps the dynamo action to occur for $R^{c}_m<17.7$. 
{ Because dynamo flows with various directions of $\bk$ may emerge spontaneously in 
a turbulent liquid, 
the presence of helicity (a consequence of the chiral anomaly) would generically 
aid the dynamo bootstrap in any geometry of the flow.} 

In conclusion, this paper proposes hydrodynamic Weyl semimetals as a host to electronic turbulence and/or dynamo effect. We derived the Navier-Stokes equations (\ref{NS}) and equations of magnetohydrodynamics (\ref{Eq:dyn-helical}) and estimated two key figures of merit -- the hydrodynamic and magnetic Reynolds numbers. Fig.~1 summarises our findings and shows that both turbulence and dynamo mechanism are in principle experimentally achievable. However, many interesting questions remain, such as experimental signatures of the turbulent electronic motion and the role of ``new'' terms in the Navier-Stokes equations, descending from the quantum chiral anomaly. Finally, we mention that while three-dimensional Dirac materials are indeed interesting from the perspective of realising the dynamo bootstrap, a number of other electronic materials may also serve as platforms to realize the effect. For example, electronic metals near critical points  (e.g., right above a superconducting transition) represent a promising system to look at in this context (both from the perspective of achieving hydrodynamic flows and  large Reynolds numbers) and could pave the way to simulating in solid-state materials the effect of magnetic field's self-excitation -- a remarkable phenomenon, usually delegated to the fields of  geophysics, astrophysics and cosmology.

{\it Acknowledgements.}
We are grateful to Matthew Foster, Anton Burkov and Aydin Keser for useful discussions.	
This research was supported by
NSF DMR-1613029 (SS),
DOE-BES (DESC0001911) (VG) and the Simons Foundation (VG).

\vspace*{-0.25in}


%


\newpage

\renewcommand{\theequation}{S\arabic{equation}}
\renewcommand{\thefigure}{S\arabic{figure}}
\renewcommand{\thetable}{S\arabic{table}}
\renewcommand{\thetable}{S\arabic{table}}
\renewcommand{\bibnumfmt}[1]{[S#1]}
\renewcommand{\citenumfont}[1]{S#1}

\setcounter{equation}{0}
\setcounter{figure}{0}
\setcounter{enumiv}{0} 

\onecolumngrid
\newpage
\begin{center}
	\textbf{\large Supplemental Material for \\
		``Dynamo Effect and Turbulence in Hydrodynamic Metals''}
\end{center}

\section{Navier-Stokes equation for a Weyl semimetal}

In this section we present a microscopic derivation of the Navier-Stokes equation (\ref{NS}) for a Weyl semimetal. In the absence
of dissipation, electromagnetic fields and internodal scattering processes, the motion of the electronic liquid in a Weyl
semimetal is Lorentz-invariant with the Fermi velocity $v_F$ playing the role of the speed of light $c$.
Indeed, so long as the crystal lattice of a Weyl semimetal is at rest, Weyl electrons propagate with velocity $v_F$
in all directions regardless of the hydrodynamic velocity $\bu$ of the electron liquid and, thus,
obey the relativistic composition law for velocities with the replacement $c\rightarrow v_F$.

The dynamics of such a liquid is described by the relativistic Navier-Stokes equation [see, for example, Ref.~\onlinecite{LL6S}].
Here, we focus on the interplay of this dynamics with the internodal scattering of electrons
(including the chiral anomaly) and electromagnetic fields, which do not transform under representations of the Lorentz group
with the speed of light replaced by the Fermi velocity.

In what follows, we set $v_F=1$ and, following Refs.~\onlinecite{LL6S} and \onlinecite{LL2S}, introduce the
four-position $x^i=(t,\br)$ and the four-velocity 
$u^i=\gamma(1,\bu)$ of the liquid. The stress-energy tensor of the Weyl electron liquid is given by
\begin{align}
	T^{ij}=\gamma w u^i u^j-P g^{ij},
	\label{T}
\end{align}
where $w$ is the enthalpy of the liquid per volume; $P$ is the pressure; $\gamma=\left(1-u^2\right)^\frac{1}{2}$; and
$g^{ij}=\text{diag}\left(1,-1,-1,-1\right)$ is the metric tensor. The equations of motion of quasiparticles
near a Weyl node
are given by
\begin{align}
	\frac{\partial T^{ij}}{\partial x^i}=\frac{1}{c}F^{jk}j_k+Q^j,
	\label{TeqMotion}
\end{align}
where $F^{ij}$ is the effective electromagnetic-field tensor, which we find below, $j_k=(\rho,-\bj)$ is the covariant four-current
of Weyl fermions near the node under consideration (in this section we suppress the node index),
with $\rho$ being the charge density (cf. Eq.~\ref{RhoMuRelation}), and
$Q^i$ is the effective ``force'' which comes from the internodal electron dynamics and which we also derive below. 
In this section, we do not consider dissipation processes due to the viscosity of the electron liquid,
as their contribution amounts to the usual viscous force in a relativistic liquid\cite{LL6S}.

{\it Lorentz force.} The contribution $S_{em}=-e\int \phi dt+\frac{e}{c}\int\bA d\br$ of
the electromagnetic field to the action of a Weyl electron corresponds to the effective electromagnetic four-potential
 $\tilde{A}^i=(c\phi,\bA)$,
defined as $S_{em}=-\frac{e}{c}\int\tilde{A}^i dx_i$, where $c$ is the speed of light
(measured in units of the Fermi velocity $v_F$). Using the corresponding electromagnetic-field tensor
\begin{align}
	F^{ij}
	\equiv \frac{\partial \tilde{A}^j}{\partial x_i}- \frac{\partial \tilde{A}^i}{\partial x_j}
	=\left(
	\begin{array}{cccc}
		0 & -cE_x & -cE_y & -cE_z \\
		cE_x & 0 & -B_z & B_y \\
		cE_y & B_z & 0 & -B_x \\
		cE_z & -B_y & B_x & 0
	\end{array}
	\right),
\end{align}
we recover the conventional expression $\frac{1}{c}F^{jk}j_k=\rho \bE+\frac{1}{c}\bj\times \bB$ for the Lorenz force
acting on a unit volume of the Weyl liquid. 

{\it Chiral anomaly and internodal scattering.} Simultaneous application of parallel electric and magnetic
fields in a Weyl semimetal results in the pumping of charge carriers from one node to the other. Impurities and
interactions may also lead to internodal scattering. All these internodal processes contribute to the time derivatives
$\frac{\partial T^{0i}}{\partial t}$ (where $i=0,x,y,z$), which corresponds to the 
four-force [cf. Eq.~(\ref{TeqMotion})] 
\begin{equation}
	Q^i=(\gamma q_w -q_P, \gamma q_w\bu),
\end{equation}
where
\begin{align}
	q_w=\frac{4}{3}\left(\frac{\partial\varepsilon}{\partial \rho}\right)
	\left(\chi\frac{ge^3}{h^2c}\bE\cdot\bB-\frac{\rho-\tilde{\rho}}{\tau_{\text{in}}}\right)
	\label{qw}
	\\
	q_P=\frac{1}{3}\left(\frac{\partial\varepsilon}{\partial \rho}\right)
	\left(\chi\frac{ge^3}{h^2c}\bE\cdot\bB-\frac{\rho-\tilde{\rho}}{\tau_{\text{in}}}\right)
	\label{qP}
\end{align}
are the the rates of change of the enthalpy and the pressure of charge carriers near a given Weyl node
due to internodal processes; $\varepsilon$ and $\chi=\pm1$ are
the internal energy (per volume) and the chirality of Weyl fermions near this node;
$\tilde{\rho}$ is the charge density at the other node. In Eqs.~(\ref{qw}) and (\ref{qP})
we have used that for Weyl fermions $P\approx \varepsilon/3$ and $w\approx 4\varepsilon/3$. The expression
$\chi\frac{ge^3}{h^2c}\bE\cdot\bB-\frac{\rho-\tilde{\rho}}{\tau_{\text{in}}}$
in Eqs.~(\ref{qw}) and (\ref{qP}) describes the change of the charge $\rho$ density near a node
due to the chiral magnetic effect and internodal scattering [cf. Eq.~(\ref{Currents})].

The change of the internal energy $\varepsilon$ when changing the number of particles near a node depends
on the heat transfer between the respective electrons and the environment. In this paper, we focus on
isothermal flows, with the system being kept at a constant temperature $T$, and assume that 
thermal equilibration near the nodes (e.g. due a phonon bath which may flow together with the electron liquid)
 takes place significantly faster than the internodal particle-transfer processes.  
Under these conditions, we find from Eqs.~(\ref{RhoMuRelation}) and (\ref{EpsilonMuRelation})
\begin{equation}
	\left(\frac{\partial\varepsilon}{\partial\rho}\right)=
	\left(\frac{\partial\varepsilon}{\partial\rho}\right)_T=\frac{3\mu}{e}\frac{\mu^2+\pi^2 T^2}{3\mu^2+\pi^2 T^2}.
\end{equation}
We emphasise, however, that the rate $\left(\frac{\partial\varepsilon}{\partial\rho}\right)$ may, in general, 
be different under different assumptions about the nature of equilibration processes. For example,
if the internodal dynamics is fast compared to internodal equilibration, the former will result in different
temperatures or even non-equilibrium distributions of electrons near different nodes.

{\it Navier-Stokes equation.} In order to obtain the Navier-Stokes equation, we consider the projection of
Eq.~(\ref{TeqMotion}) on the direction perpendicular to the four-velocity $u^i$:
\begin{align}
	\frac{\partial T^{ji}}{\partial x^j}-u^iu_l\frac{\partial T^{lk}}{\partial x^l}
	=\frac{1}{c}F^{ik}j_k-\frac{1}{c}u^iu_n F^{nk}j_k
	+Q^i-u^iu_jQ^j.
	\label{Projection}
\end{align}
Considering the vector components ($i=x,y,z$) of Eq.~(\ref{Projection}) and using Eqs.~(\ref{T}), (\ref{qw}) and (\ref{qP})
and that $u_iu^i=1$, we arrive at the Navier-Stokes equation
\begin{align}
	w\left(\frac{\partial}{\partial t}+\bu\cdot\bnabla\right)\bu
	=-\bnabla P-\bu\frac{\partial P}{\partial t} +\rho\bE+\frac{1}{c}\,\bj\times \bB
	+\frac{\bu}{3}\left(\frac{\partial\varepsilon}{\partial\rho}\right)
	\left(\chi\frac{ge^3}{h^2c}\bE\cdot\bB-\frac{\rho-\tilde{\rho}}{\tau_{\text{in}}}\right),
\end{align}
where we neglected corrections of higher orders in $\bu^2$ to each term.


\section{Ponomarenko dynamo and helicity}

In this section we provide the details of derivations of equations governing the kinematic Ponomarenko dynamo for a given flow velocity. For completeness in Sec.~\ref{SI:dynamoconventional} we present the conventional Ponomarenko dynamo, and then the effects of helicity are examined in
 Sec.~\ref{SI:dynamohelicity}. 

\subsection{Conventional Ponomarenko Dynamo \label{SI:dynamoconventional}}
The dynamo equation reads as 

\bea \frac{\partial\mathbf{B}}{\partial t}=\mathbf{B}\cdot\boldsymbol{\nabla}\mathbf{u}-\mathbf{u}\cdot\boldsymbol{\nabla}\mathbf{B}+\frac{c^2}{4\pi\sigma} \nabla^2\mathbf{B}. \eea

{ Following Ref.~[\onlinecite{plasmaS}],} we assume that a cylinder is filled up with a conductive liquid with a flow field as

\[ \mathbf{u} =
\begin{cases}
(0,r\Omega, u_0)  & r\leq a\\
0  & r\geq a
\end{cases},
\]
where $a$ is the radius of the cylinder, $\Omega$ is the angular velocity and $u_{0}$ is the velocity along the axis, all taken to be constant. Assuming the ansatz  
\bea \mathbf{B}(r,\theta,z,t)=\mathbf{B}(r) e^{i(n\theta-kz)+\gamma t}  \eea
for the magnetic field and plugging into the dynamo equation, we obtain the $r$- and $\theta$- components of the magnetic fields as 

\bea \label{Br_stan} y^2 \frac{d^2B_r}{dy^2}+y \frac{dB_r}{dy}=(q^2y^2+n^2+1)B_r+2inB_{\theta},\\  \label{Bt_stan}
y^2 \frac{d^2B_\theta}{dy^2}+y \frac{dB_\theta}{dy}=(q^2y^2+n^2+1)B_\theta-2inB_r,
\eea   

The $z$-component of the magnetic field doesn't enter the equations above. It can be determined from $\boldsymbol{\nabla}\cdot\mathbf{B}=0$. Letting $B_{\pm}=B_r\pm B_{\theta}$, we get a set of separable equations

\bea \label{Bpm_stan1} y^2 \frac{d^2B_{\pm}}{dy^2}+y \frac{dB_{\pm}}{dy}-[q^2y^2+(n\pm1)^2]B_{\pm}=0,~~~y\leq a, \\ \label{Bpm_stan2}
y^2 \frac{d^2B_{\pm}}{dy^2}+y \frac{dB_{\pm}}{dy}-[s^2y^2+(n\pm1)^2]B_{\pm}=0,~~~y> a,
\eea
where 
\bea y=\frac{r}{a},~~ \tau_{R}=\frac{4\pi\sigma a^2}{c^2},~~q^2=k^2a^2+\gamma\tau_{R}+i(n\Omega-ku_{0})\tau_{R},~~s^2=k^2a^2+\gamma\tau_{R}.  \eea

Equations (\ref{Bpm_stan1}-\ref{Bpm_stan2}) are modified Bessel equations with the solutions 
\bea \label{regBsl_stan} B_{\pm}(y)=C_{\pm}\frac{I_{n\pm1}(qy)}{I_{n\pm1}(q)},~~~y\leq 1,\\ \label{sinBsl_stan}
B_{\pm}(y)=D_{\pm}\frac{K_{n\pm1}(sy)}{K_{n\pm1}(s)},~~~y> 1,
\eea
where $I_{n}$ and $K_{n}$ are modified Bessel functions. Continuity of the fields across the boundary yields $C_{\pm}=D_{\pm}$. The second matching condition is given due the jumping of angular velocity across the boundary: 

\bea \left[\frac{dB_{\pm}}{dy}\right]^{y=1_{+}}_{y=1_{-}}=\pm i\Omega\tau_{R}\left(\frac{B_{+}+B_{-}}{2}\right).\eea

Inserting the solutions (\ref{regBsl_stan}) and (\ref{sinBsl_stan}) in the equations above, we obtain the dispersion relation as 
\bea \label{disp_stan} G_{+}G_{-}=\frac{i}{2}\Omega\tau_{R}(G_{+}-G_{-}), \eea
where 
\bea G_{\pm}=q\frac{I'_{n\pm1}(q)}{I_{n\pm1}(q)}-s\frac{K'_{n\pm1}(s)}{K_{n\pm1}(s)}. \eea 
Here, $'$ denotes derivative with respect to the argument. 



Our aim would be to obtain the magnetic Reynold number $R_{m}=\tau_{R}/\tau_{H}$, where $\tau_{H}=a/v$. Here $v$ denotes the typical velocity of the flow. Taking $v=\sqrt{\Omega^2 a^2+u^2_{0}}$, the $R_{m}$ reads as 
\bea \label{Reyld_mag} R_{m}=\frac{\tau_{R}\sqrt{\Omega^2 a^2+u^2_{0}}}{a}.\eea 

To find the critical magnetic Reynold number $R^c_m$, beyond which the dynamo action commences, we have to solve equation (\ref{disp_stan}) numerically. We set $\mathrm{Re}(\gamma)=0$, the onset value beyond which the magnetic field grows exponentially for $\mathrm{Re}(\gamma)>0$. We vary $ka$ and $\mathrm{Im}(\gamma \tau_R)$ over a wide range of values and look for unknown variables $\Omega\tau_R$ and $u_0\tau_{R}/a$, all dimensionless, through the imaginary and real parts of the dispersion relation (\ref{disp_stan}). The $n=1$ is the first dynamo mode excited with $R^{c}_{m}\simeq 17.72$.

\subsection{Ponomarenko dynamo with helicity term\label{SI:dynamohelicity}}
Now we add the helical term to the dynamo equation as

\bea \frac{\partial\mathbf{B}}{\partial t}=\mathbf{B}\cdot\boldsymbol{\nabla}\mathbf{u}-\mathbf{u}\cdot\boldsymbol{\nabla}\mathbf{B}+\frac{c^2}{4\pi\sigma} \nabla^2\mathbf{B}+\xi \boldsymbol{\nabla}\times \mathbf{B}.\eea

In writing down the last term we assumed that the $\xi$ is constant to simplify the subsequent equations. In principle it does depend on the magnetic and electric fields. The last term $\boldsymbol{\nabla}\times \mathbf{B}$ gives rise to new terms involving $B_z$:

\bea \left[\frac{in}{r}B_z(r)+ikB_{\theta}(r) \right]e^{i(n\theta-kz)+\gamma t} \eea
in the $r$-component and 

\bea \left[-ikB_r(r)-\frac{\partial{B_{z}(r)}}{\partial r} \right]e^{i(n\theta-kz)+\gamma t} \eea
in the $\theta$-component, which are added to right hand side of Eqs. (\ref{Br_stan}) and (\ref{Bt_stan}), respectively. 

Using $\boldsymbol{\nabla}\cdot\mathbf{B}=0$, we write the $z$-component as 
\bea B_{z}(r)=\frac{1}{ikr}B_{r}(r)+\frac{1}{ik}\frac{dB_r(r)}{dr}+\frac{n}{kr}B_{\theta}(r) \eea

\bea \label{Br_chi} y^2 \frac{d^2B_r}{dy^2}+y \frac{dB_r}{dy}&=&(q^2y^2+n^2+1)B_r+2inB_{\theta}-\delta \left(nB_r+ny\frac{dB_r}{dy}+in^2B_{\theta}+ik^2a^2y^2B_{\theta}  \right)  ,\\ \nonumber
y^2 \frac{d^2B_\theta}{dy^2}+y \frac{dB_\theta}{dy}&=&(q^2y^2+n^2+1)B_\theta-2inB_r-\delta \left(-ik^2a^2y^2B_{r}-iB_{r}+iy\frac{dB_{r}}{dy}+iy^2\frac{d^2B_{r}}{dy^2}+nB_{\theta}-ny\frac{dB_{\theta}}{dy} \right)\\ \label{Bt_chi}
\eea 
where $\delta=\tau_R/\tau_c$ with $\tau_c=ka^2/\xi$. In the limit $\xi=0$ ($\delta\rightarrow0$) the equations above reduce to Eqs. (\ref{Br_stan}) and (\ref{Bt_stan}). In (\ref{Bt_chi}) we replace the $y^2\frac{d^2B_{r}}{dy^2}+y\frac{dB_{r}}{dy}$ in the parentheses with the expression (\ref{Br_chi}) and keep the terms up to first order in $\delta$. We get 

\bea y^2 \frac{d^2B_r}{dy^2}+y \frac{dB_r}{dy}&=&(q^2y^2+n^2+1)B_r+2inB_{\theta}-\delta \left(nB_r+ny\frac{dB_r}{dy}+in^2B_{\theta}+ik^2a^2y^2B_{\theta}  \right)  ,\\ 
y^2 \frac{d^2B_\theta}{dy^2}+y \frac{dB_\theta}{dy}&=&(q^2y^2+n^2+1)B_\theta-2inB_r-\delta\left(-ik^2a^2y^2B_{r}+i(q^2y^2+n^2)B_r-nB_{\theta}-ny\frac{dB_{\theta}}{dy}    \right) 
\eea 

Rewriting the equations above in terms of $B_{\pm}=B_r\pm iB_{\theta}$, we obtain  

\bea  \label{Bp} y^2 \frac{d^2B_{+}}{dy^2}+y \frac{dB_{+}}{dy}&=&\left[q^2y^2+(n+1)^2\right]B_{+}-\delta\left(ny\frac{dB_{-}}{dy}-n(n-1)B_{-}+k^2a^2y^2B_{+}-q^2y^2B_r  \right)  ,\\
y^2 \frac{d^2B_{-}}{dy^2}+y \frac{dB_{-}}{dy}&=&\left[q^2y^2+(n-1)^2\right]B_{-}-\delta\left(ny\frac{dB_{+}}{dy}+n(n+1)B_{+}-k^2a^2y^2B_{-}+q^2y^2B_r   \right). \label{Bm}
\eea 

In order to examine the effect of helicity $\delta$ on the magnetic Reynold number, in what follows we use equations (\ref{Bp}) and (\ref{Bm}) to evaluate the $R^{c}_{m}$ discussed in the preceding subsection. 

We rewrite the equations as 

\bea  \label{Bp_m} y^2 \frac{d^2B_{+}}{dy^2}+y \frac{dB_{+}}{dy}&=&\left[q_{+}^2y^2+(n+1)^2\right]B_{+}-\delta\left(ny\frac{dB_{-}}{dy}-n(n-1)B_{-}-\frac{q^2y^2}{2}B_{-}  \right)  ,\\ \label{Bm_m}
y^2 \frac{d^2B_{-}}{dy^2}+y \frac{dB_{-}}{dy}&=&\left[q_{-}^2y^2+(n-1)^2\right]B_{-}-\delta\left(ny\frac{dB_{+}}{dy}+n(n+1)B_{+}+\frac{q^2y^2}{2}B_{+} \right),
\eea 
where 
\bea \label{qpm}
q_{+}^2=\left(1+\frac{\delta}{2}\right)q^2-\delta k^2a^2,~~~q_{-}^2=\left(1-\frac{\delta}{2}\right)q^2+\delta k^2a^2.
\eea

Without the terms in the parentheses in the right side, we have a set of homogenous equations with helicity included. 
\bea  y^2 \frac{d^2B_{0,+}}{dy^2}+y \frac{dB_{0,+}}{dy}&=&\left[q_{+}^2y^2+(n+1)^2\right]B_{0,+},\\ 
y^2 \frac{d^2B_{0,-}}{dy^2}+y \frac{dB_{0,-}}{dy}&=&\left[q_{-}^2y^2+(n-1)^2\right]B_{0,-}.\eea 

Theses are modified Bessel equations with solutions 
\bea B_{0,\pm}(y)\propto I_{n\pm1}(q_{\pm}y), K_{n\pm1}(q_{\pm}y).
\eea

Using the above homogenous solutions, we use the perturbation theory to solve the equations in (\ref{Bp_m}-\ref{Bm_m}). We write the solutions up to the first order in $\delta$ as 
\bea \label{ansatz1} B_{+}(y)=B_{0,+}(y)+\delta A_{+}(y),~~~B_{-}(y)=B_{0,-}(y)+\delta A_{-}(y),\eea
Plugging (\ref{ansatz1}) into the equations (\ref{Bp_m}-\ref{Bm_m}), we obtain 

\bea  y^2 \frac{d^2A_{+}}{dy^2}+y \frac{dA_{+}}{dy}&=&\left[q_{+}^2y^2+(n+1)^2\right]A_{+}-\left(ny\frac{dB_{0,-}}{dy}-n(n-1)B_{0,-}-\frac{q^2y^2}{2}B_{0,-}  \right)  ,\\ 
y^2 \frac{d^2A_{-}}{dy^2}+y \frac{dA_{-}}{dy}&=&\left[q_{-}^2y^2+(n-1)^2\right]A_{-}-\left(ny\frac{dB_{0,+}}{dy}+n(n+1)B_{0,+}+\frac{q^2y^2}{2}B_{0,+} \right).
\eea 

We set $n=1$. Using the standard approaches for solving the non-homogeneous differential equations, and after a lengthy but straightforward calculations, we obtain 
\bea &&A_{+}(y)\simeq \left( \frac{q^2-q^2_{-}}{12}y^4+\frac{  3q^2q_{-}^2+q_{+}^2(q^2-q^2_{-}) }{192}y^6 \right)\\
&&A_{-}(y)\simeq-\frac{q_{+}^2}{32} y^4. 
\eea

Matching the fields across the boundary at $y=1$, we obtain a dispersion relation, which reads as 
\bea \label{disp_pr} G_{+}+\delta G\left(A^{y=1}_{+}s\frac{K'_{2}(s)}{K_{2}(s)}-A'^{y=1}_{+} \right)=-i\Omega\tau_{R}\left(1-G+\delta A^{y=1}_{-}-\delta G A^{y=1}_{+}\right), \eea
where 
\bea G=\frac{G_{+}-\delta\left( A^{y=1}_{-}s\frac{K'_{0}(s)}{K_{0}(s)}-A'^{y=1}_{-} \right)  }{ G_{-}-\delta\left( A^{y=1}_{+}s\frac{K'_{2}(s)}{K_{2}(s)}-A'^{y=1}_{+} \right) }. \eea

Again we solve the dispersion relation (\ref{disp_pr}) numerically to obtain $R_{m}^{c}$; the results are shown in Fig.~2 in the main text.


\end{document}